\documentclass[jgrga]{./HughesTeX}

\usepackage{graphicx}
\usepackage{caption}
\usepackage{amsmath}

\authorrunninghead{HUGHES, J.~M.}

\titlerunninghead{Angular Momentum Misconceptions}

\authoraddr{Corresponding author: John Hughes, Department of Physical Sciences, Embry-Riddle Aeronautical University, Daytona Beach, FL, USA. (john.hughes@erau.edu)}

\let\vec\mathbf
\usepackage{bm}
\newcommand{\vect}[1]{\boldsymbol{\mathbf{#1}}}

\begin{document}


\title{Common Misconceptions About Angular Momentum}




\authors{J.~M.~Hughes\altaffilmark{1}}

\altaffiltext{1}{Department of Physical Sciences, Embry-Riddle Aeronautical University, Daytona Beach, FL, USA.}




\begin{abstract}

Angular momentum is taught in every classical mechanics course. It is a difficult topic with misconceptions commonly forming significant barriers to student success. My intention in writing this paper is to combat some of the most common misconceptions and to help prepare freshman and sophomore students for success in their upper-level classical mechanics courses. The paper begins with a discussion of several fundamental equations relating to angular momentum. A number of special cases are then examined and the paper concludes after presenting a set of three examples.

\end{abstract}



\begin{article}

\section{Introduction}

Given a three dimensional coordinate system and a momentum $\vec{p}=m\vec{v}$, angular momentum $\vec{L}$ in classical mechanics is the ``moment of momentum'' defined by $\vec{L}=\vec{r}\times \vec{p}$ where $m$ is an element of mass, $\vec{r}$ is its position vector from some chosen origin, and $\vec{v}$ is the time derivative $d\vec{r}/dt$. This article discusses common misconceptions about angular momentum and is intended for freshman and sophomores after completion of a first course in physics where angular momentum has been introduced. I'll therefore assume familiarity with the common variables and definitions such as $\vec{a}=d\vec{v}/dt$ for acceleration, $\vect{\alpha}=d\vect{\omega}/dt$ for angular acceleration, $\vec{F}$ for force, $\vect{\tau}$ for torque, and so on. Sometimes I'll use the convenient dot notation for time derivatives where $d\left(\;\right)/dt \equiv \dot{\left(\;\right)}$ so that, for example, $\vec{a}=\dot{\vec{v}}=\ddot{\vec{r}}$.

My intention with this paper is to help prepare students for success in upper-level classical mechanics courses by combating a few of the most common misconceptions about angular momentum as typically taught in freshman physics courses. After developing and discussing some fundamental points about angular momentum, the paper ends with a selection of examples.

\section{Angular Momentum and Its Time Derivative}

Freshman physics students are often taught in analogy with $\vec{p}=m\vec{v}$  that $\vec{L}=I\vect{\omega}$ (where $I$ and $\vect{\omega}$ are the the scalar moment of inertia and vector angular velocity, respectively) and in analogy with Newton's second law that $\sum\vect{\tau}=d\vec{L}/dt=I\vect{\alpha}$. Unfortunately, none of those angular equations are correct in general. Also unfortunate are the misconceptions they implant in students' minds and the obstacles they form to the proper understanding required for success in upper-level classical mechanics courses. This section gives and discusses the correct form of the relevant equations and the subsequent section examines the conditions under which the special cases $\vec{L}=I\vect{\omega}$, $\sum\vect{\tau}=d\vec{L}/dt$, and $\sum\vect{\tau}=I\vect{\alpha}$ are true.

\subsection{Angular Momentum of An Object in General Motion}

\begin{center}
\begin{figure}[h]
\noindent\includegraphics[width=18pc]{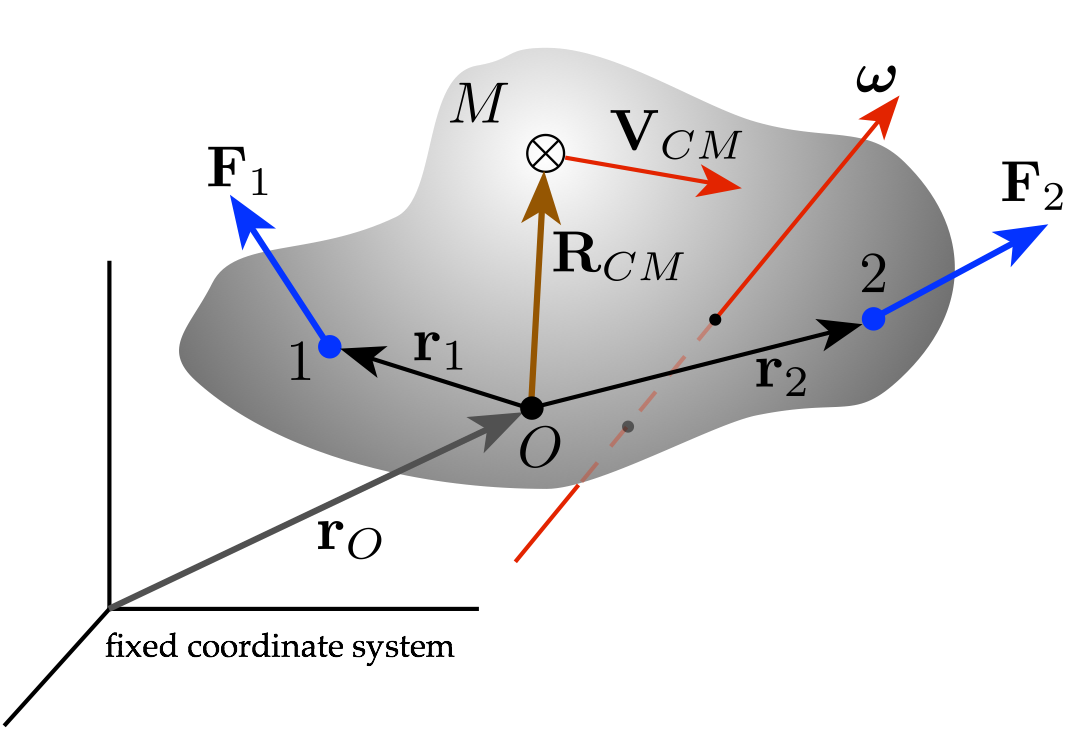}
\caption{An arbitrary three-dimensional object in general motion, both translating and rotating.}
\label{fig:3D_object}
\end{figure}
\end{center}

Consider the arbritrary three-dimensional object of mass $M$ in Figure \ref{fig:3D_object} and suppose it is in general motion, both translating as a whole with some velocity relative to the fixed coordinate system and rotating with angular velocity $\vect{\omega}$ about some arbitrary axis. External forces $\vec{F}_i$ where $i=1,2,3,...$ act on the object at the arbitrary positions $\vec{r}_i$. The object's center of mass ($CM$) is labeled $\otimes$ and located at $\vec{R}_{CM}=\int \vec{r}\,dm/M$ relative to origin point $O$.

The object's angular momentum relative to origin point $O$ is obtained by integrating $d\vec{L}=\vec{r}\times d\vec{p}=\vec{r}\times\vec{v}dm$ over the object's volume, where all vectors are relative to $O$ so that $\vec{r}=\vec{r}'-\vec{r}_O$ and $d\vec{p}=\left(\dot{\vec{r}'}-\dot{\vec{r}}_O\right)dm$ with the prime ($'$) on vectors other than $\vec{r}_O$ indicating measurement relative to the fixed coordinate system. The necessary integral and the resulting algebra are best left for an upper-level classical mechanics course, but the result can be expressed as
\begin{equation}
\vec{L}=M\left(\vec{R}_{CM}\times\vec{V}_{CM}\right)+\vec{L}_{{\mathrm{rot}}_{CM}}
\label{eq:general_L}
\end{equation}
where $\vec{V}_{CM}=\dot{\vec{R}}_{CM}$ is the instantaneous center of mass velocity and $\vec{L}_{\mathrm{rot}_{CM}}$ is the object's angular momentum from rotation about its $CM$. Equation \ref{eq:general_L} reflects the useful truth than any rigid object's general motion can be represented by translation of the whole without rotation plus rotation of the whole without translation. In this expression, the translation is at the center of mass velocity and the rotation is about an axis through the center of mass. We will return to this useful equation after examining the time derivative of $\vec{L}$.

\subsection{The Relation of Torque to $d\vec{L}/dt$}

As above, the angular momentum about $O$ of the object in Figure \ref{fig:3D_object} is obtained by evaluating the integral 
\begin{displaymath}
\vec{L}=\int \vec{r}\times d\vec{p}=\int\left(\vec{r}'-\vec{r}_O\right)\times\left(\dot{\vec{r}'}-\dot{\vec{r}}_O\right)dm
\end{displaymath}
over the object's volume. Again leaving the derivation details for an upper-level course, taking the time derivative of this equation and performing some algebra gives
\begin{equation}
\frac{d\vec{L}}{dt}=\sum\vect{\tau}-M\left(\vec{R}_{CM}-\vec{r}_O\right)\times\ddot{\vec{r}}_O
\label{eq:L_dot}
\end{equation}
where $\sum\vect{\tau}=\sum\left(\vec{r}_i\times\vec{F}_i\right)$ is the net external torque about $O$ due to the forces $\vec{F}_i$. This interesting equation, of much use in classical mechanics, is a correct general relation between torque and $d\vec{L}/dt$.

\subsection{The Inertia Tensor}

At this point we need a brief aside to introduce the inertia tensor. Equation \ref{eq:general_L} nicely separates the contributions from translation and rotation to an object's angular momentum. Considering a purely rotating object so that $\vec{v}=\vect{\omega}\times\vec{r}$, substitution into $d\vec{L}_{\mathrm{rot}}=\vec{r}\times\vec{v}\,dm$ and integration over the object's mass results in 
\begin{equation}
\vec{L}_{\mathrm{rot}}=\vec{I}\cdot\vect{\omega}
\end{equation}
where $\vec{I}$ is the inertia tensor given by
\begin{displaymath}
\begin{array}{lcl}
 \vec{I} & = &
  \left( {\begin{array}{ccc}
   I_{xx} & I_{xy} & I_{xz} \\ I_{yx} & I_{yy} & I_{yz} \\ I_{zx} & I_{zy} & I_{zz}
  \end{array} } \right)\\
 & = & \left( {\begin{array}{ccc}
   \int(y^2+z^2) & -\int(xy) & -\int(xz) \\ -\int(xy) & \int(z^2+x^2) & -\int(yz) \\ -\int(xz) & -\int(yz) & \int(x^2+y^2)
  \end{array} } \right)dm.
  \end{array}
\end{displaymath}

The inertia tensor is a 3$\times$3 matrix with six independent elements calculated from the object's mass distribution. Asymmetries in the mass distribution give rise to off-diagonal elements and consequent components of $\vec{L}_{{\mathrm{rot}}}$ not parallel to those of $\vect{\omega}$. Any off-diagonal elements vanish when $\vec{I}$ is calculated for the so-called principal axes. For example, an axis of symmetry is a principal axis.

\section{Special Cases}
\subsection{Obtaining $\vec{L}=I\vect{\omega}$ \label{sec:LIw}}

Repeating equation \ref{eq:general_L}, an object's angular momentum is in general given by $\vec{L}=M\left(\vec{R}_{CM}\times\vec{V}_{CM}\right)+\vec{L}_{\mathrm{rot}_{CM}}$. The most common situations for which this reduces to $\vec{L}=I\vect{\omega}$, where $I$ is a scalar, require three special conditions: 
\begin{itemize}
\item[1.] The object is in purely rotational motion about its $CM$ with no translation. Equation \ref{eq:general_L} then reduces to $\vec{L}=\vec{L}_{\mathrm{rot}_{CM}}$, the object's angular momentum from rotation about its $CM$. 
\item[2.] The coordinate system in which $\vec{L}$ is measured is coincident with the object's principal axes (for example, axes of symmetry) so that the inertia tensor has no off-diagonal elements. 
\item[3.] The object is rotating about one of its principal axes. 
\end{itemize}

With these three conditions holding, let $\vect{\omega}=\omega\hat{z}$ and $I_{zz}=I$ so that
\begin{displaymath}
\vec{L}=\vec{L}_{CM_{\mathrm{rot}}}=
\vec{I}\cdot\vect{\omega}=
\left( {\begin{array}{ccc}
I_{xx} & 0 & 0 \\ 0 & I_{yy} & 0 \\ 0 & 0 & I
\end{array} } \right)
\left( {\begin{array}{c}
0 \\ 0 \\ \omega 
\end{array}}\right)
= I \omega \hat{z}
\end{displaymath}
and $\vec{L}=I\vect{\omega}$. Students should carefully note that $\vec{L}$ and $\vect{\omega}$ are generally {\em not} parallel to each other, although they are in this special case of a pure rotation of a symmetrical object about an axis of symmetry.

\subsection{Obtaining $\sum\vect{\tau}=d\vec{L}/{dt}$ and $\sum\vect{\tau}=I\vect{\alpha}$}

Repeating equation \ref{eq:L_dot}, the net torque on an object is related to its change in angular momentum by 
\begin{displaymath}
\frac{d\vec{L}}{dt}=\sum\vect{\tau}-M\left(\vec{R}_{CM}-\vec{r}_O\right)\times\ddot{\vec{r}}_O. 
\end{displaymath}
where $\left(\vec{R}_{CM}-\vec{r}_O\right)$ locates the $CM$ relative to the origin point $O$ about which torques and angular momentum are calculated and the quantity $\ddot{\vec{r}}_O$ is the origin point's acceleration. There are three nontrivial ways to make $M\left(\vec{R}_{CM}-\vec{r}_O\right)\times\ddot{\vec{r}}$ vanish so that $\sum\vect{\tau}=d\vec{L}/dt$. Either
\begin{itemize}
\item[1.] $\vec{R}_{CM}-\vec{r}_O=0$ (the origin point is located at the $CM$) so that torques and angular momentum are calculated about the $CM$, or
\item[2.] $\ddot{\vec{r}}_O=0$ so the origin point (wherever it is located) is not accelerating, or
\item[3.] $\left(\vec{R}_{CM}-\vec{r}_O\right)$ is parallel to $\ddot{\vec{r}}_O$ so the origin point's acceleration is along the line joining the origin point to the $CM$ (making the cross product vanish). This relatively obscure condition is rarely invoked.
\end{itemize}

Under any of these three conditions, $\sum\vect{\tau}=d\vec{L}/dt$. 

Obtaining $\sum\vect{\tau}=I\vect{\alpha}$ requires two further conditions:
\begin{itemize}
\item[4.] $\vec{L}=I\vect{\omega}$ as in \S\ref{sec:LIw} so that $\sum\vect{\tau}=\frac{d\vec{L}}{dt}=\frac{d\left(I\vect{\omega}\right)}{dt}$.
\item[5.] $I$ is constant in time, meaning the object is rigid so that its shape never changes and $\sum\vect{\tau}=I\frac{d\vect{\omega}}{dt}$.
\end{itemize}
Only then, by the definition of angular acceleration, does $\sum\vect{\tau}=I\vect{\alpha}$.

To summarize this special case, perhaps the most common ways of setting $\sum\vect{\tau}=d\vec{L}/dt$ are to place the origin point either at (a) the $CM$ or (b) a fixed location. The further simplification of setting $\sum\vect{\tau}=I\vect{\alpha}$ requires a rigid object with $\vec{L}=I\vect{\omega}$ as in \S\ref{sec:LIw}.

\section{Three Examples}


\subsection{The Direction of $\vec{L}$}

From $\vec{p}=m\vec{v}$, momentum is always parallel to velocity, a fact that apparently contributes to the common misconception that angular momentum $\vec{L}$ is always parallel to angular velocity $\vect{\omega}$. It is not, and the following example is intended to combat this misconception.

\begin{center}
\begin{figure}[h!]
\noindent
\includegraphics[width=17pc]{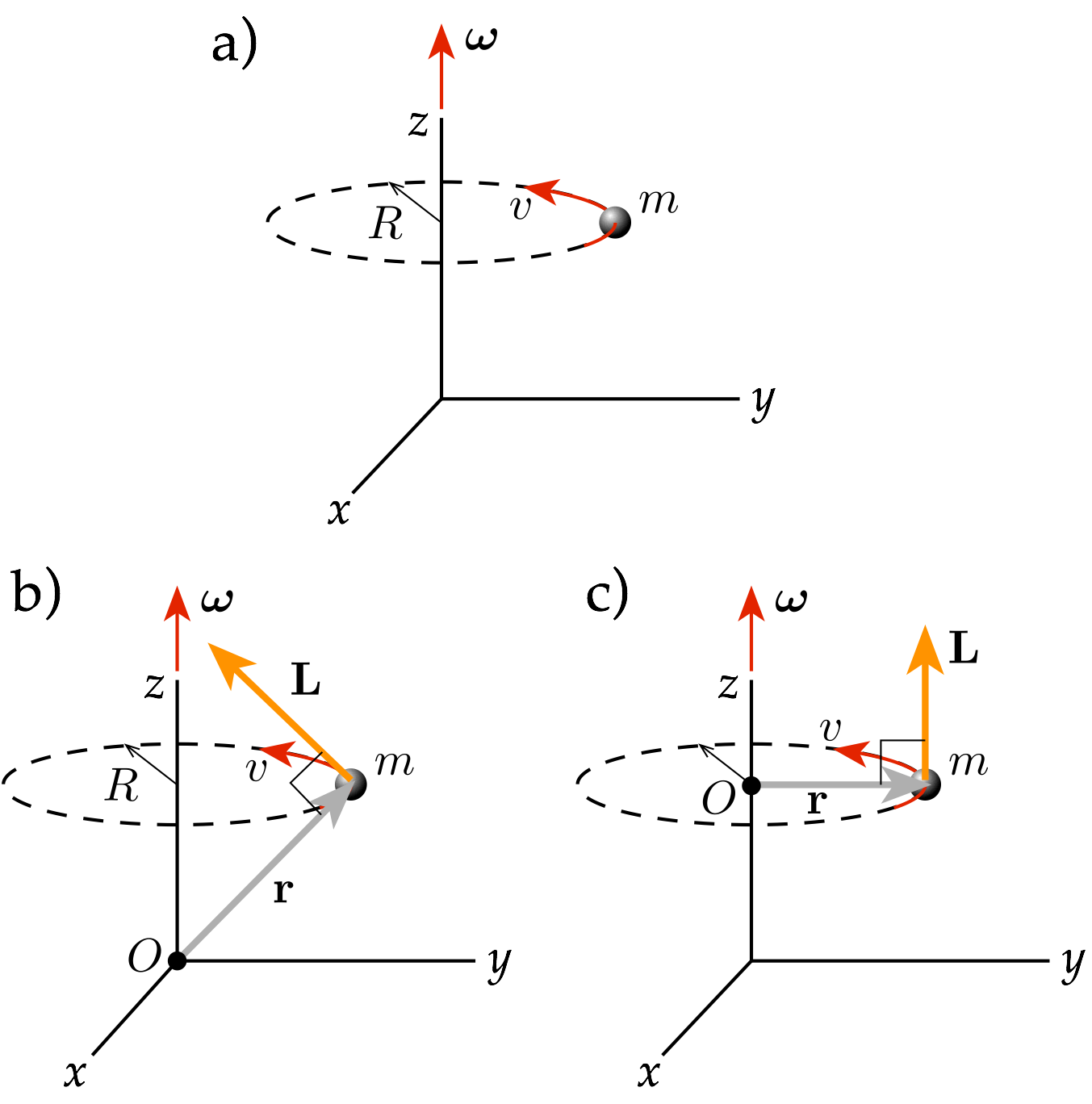}
\caption{a) A point mass in circular motion. b) The object's position and angular momentum vectors relative to the origin. c) The object's position and angular momentum vectors relative to the center of its orbit.}
\label{fig:point_particle}
\end{figure}
\end{center}

Consider the particle of mass $m$ in Figure \ref{fig:point_particle}a moving along a circular path of radius $R$ at constant speed $v$. By the right-hand-rule for angular velocity, $\vect{\omega}$ is in the $+\hat{z}$ direction so that $\vect{\omega}=\omega\hat{z}$. Its angular momentum is, by definition, $\vec{L}=\vec{r}\times\vec{p}=m\left(\vec{r}\times\vec{v}\right)$ and cannot be evaluated before an origin point is specified. The vector $\vec{r}$ then extends from the origin point to the particle and $\vec{L}$ is the angular momentum relative to that chosen point. Let us choose two origin points and evaluate the direction of $\vec{L}$ in each case. 

First, chose the origin point $O$ as in Figure \ref{fig:point_particle}b. Applying the right-hand-rule to determine the direction of $\vec{r}\times\vec{v}$ shows that $\vec{L}$ has components along $-\hat{y}$ and $+\hat{z}$ as shown and is clearly not parallel to $\vect{\omega}$.

Next, place the origin point $O$ at the circular path's center as in Figure \ref{fig:point_particle}c so that $\vec{r}$ lies in the particle's orbital plane. In this case, applying the right-hand-rule as before shows that $\vec{L}=L\hat{z}$ so that it is parallel to $\vect{\omega}$.

So is $\vec{L}$ parallel to $\vect{\omega}$, or is it not? The question is ill-posed until an origin point is specified. Because angular momentum can be computed about any origin point, a particular momentum $\vec{p}$ can have an angular momentum $\vec{L}$ of any magnitude in any direction. Students are encouraged to remember this: angular momentum is the moment of momentum; for a mass element $dm$, it is it is {\it always} in the direction of $d\vec{L}=\vec{r}\times d\vec{p}=\vec{r}\times \vec{v}\,dm$.

\subsection{The Magnitude of $\vec{L}$}

\begin{center}
\begin{figure}[h!]
\noindent\includegraphics[width=18pc]{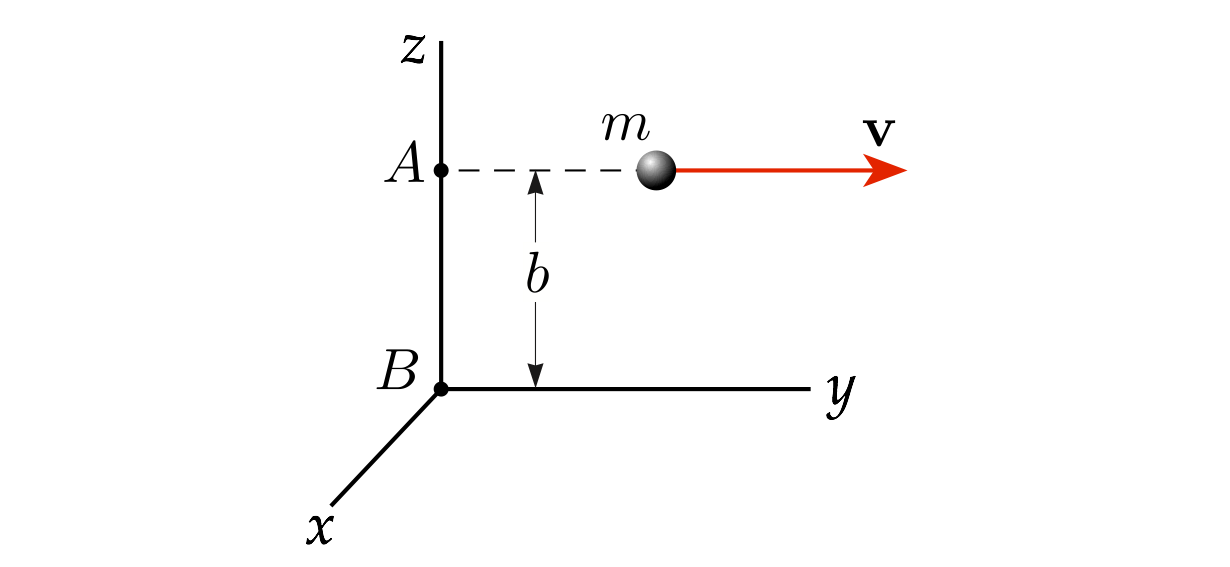}
\caption{A point mass moving with constant velocity.}
\label{fig:point_particle2}
\end{figure}
\end{center}

Another common misconception relating to $\vec{L}$ is the mistaken view that only rotating objects have angular momentum. The particle in Figure \ref{fig:point_particle2} is moving in a straight line at constant speed; its velocity is constant and it is not rotating about any axis through itself. Following the arguments of the previous example, its angular momentum relative to origin point $A$ equals zero (because $\vec{r}$ extending from $A$ to $m$ is parallel to $\vec{v}$ so the cross product vanishes) while its angular momentum relative to origin point $B$ is not zero. In this latter case, $\vec{L}=-mvb\hat{x}$.

In general, an object does not need to be rotating to have angular momentum. It has angular momentum whenever it has a moment of momentum, and this equals zero for a point particle with $\vec{L}=\vec{r}\times\vec{p}$ only when $\vec{r}=0$, $\vec{p}=0$, or when $\vec{r}$ is parallel to $\vec{p}$.

\subsection{A More Difficult Example}

The following problem is from \citet{Morin} Exercise 9.53. It can be solved using only $\sum\vec{F}=m\vec{a}$ (try it!) but here it is solved using angular momentum and torque, mostly because that is the topic of this paper, but partly because it is fun.

\begin{center}
\begin{figure}[h!]
\noindent\includegraphics[width=18pc]{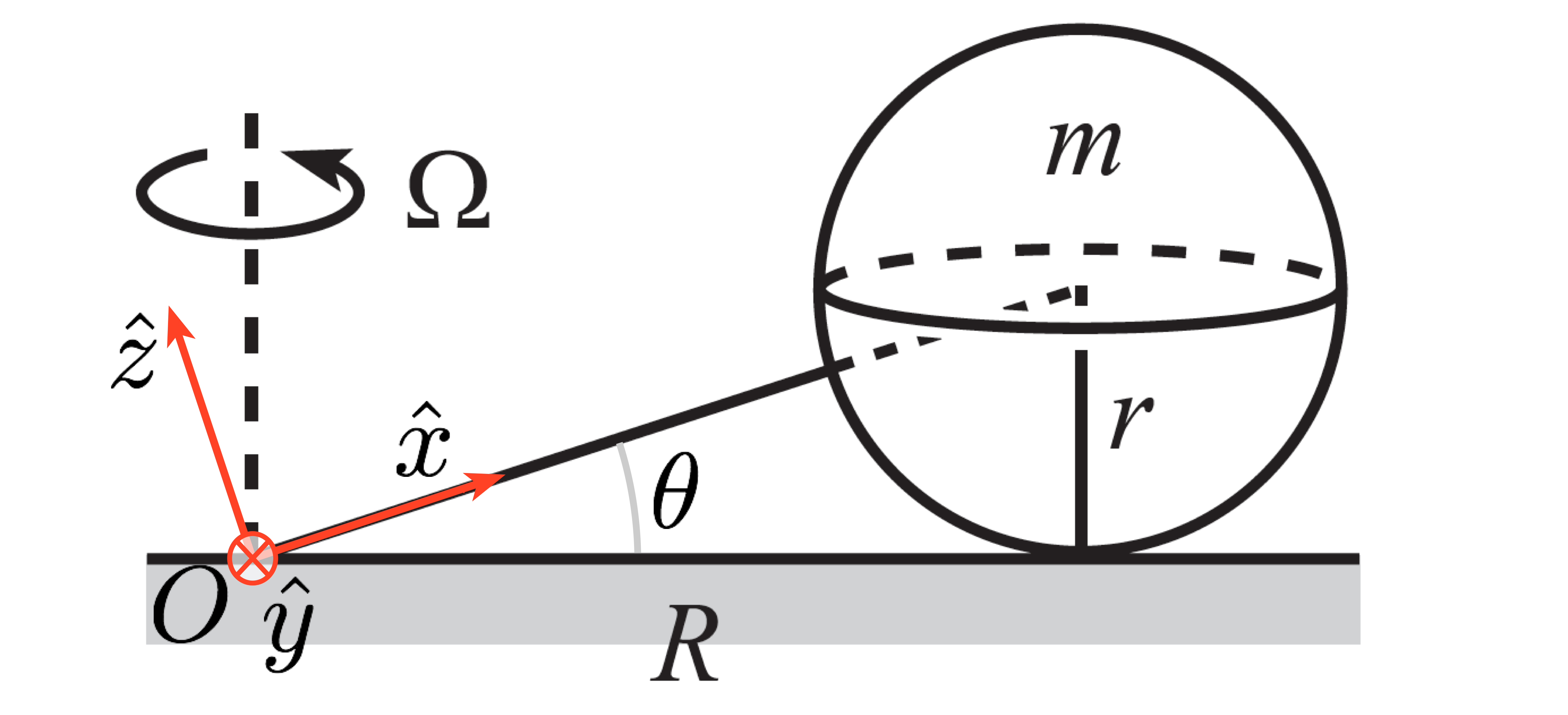}
\caption{A lollipop on a massless stick, sliding on frictionless ground around a circle of radius $R$ at angular speed $\Omega$. (Modified from \citep{Morin} Fig 9.67.)}
\label{fig:09_53}
\end{figure}
\end{center}

\vspace{0.2cm}
\noindent Problem:

\vspace{0.2cm}
\noindent Consider the lollipop in Figure \ref{fig:09_53} made of a solid sphere of mass $m$ and radius $r$ that is radially pierced by a massless stick. The free end of the stick is pivoted on frictionless ground. The sphere $slides$ along the ground, with the same point on the sphere always touching the ground. The sphere's center moves in a circle of radius $R$ with frequency $\Omega$. Using $\sum\vect{\tau}=d\vec{L}/dt$, show that the normal force between the ground and the sphere is $N = mg + mr\Omega^2$.

\vspace{0.2cm}

\noindent Solution:

\vspace{0.2cm}
We will solve this problem in the following steps. First, an origin point and coordinate system will be chosen so that $\vec{L}$ can be evaluated. Next, we'll use the fact that the lollipop is rotating at frequency $\Omega$ to evaluate the time derivative $d\vec{L}/dt$. Last, equating that result with the net toque on the lollipop will result in an expression that can be solved for the required normal force.

\vspace{0.4cm}
\noindent{{Choosing the Origin Point and Coordinate System:}}
\vspace{0.2cm}

Three external forces act on the lollipop: the lollipop's weight, the normal force (which keeps the sphere from falling through the frictionless ground) and tension on the lower left end of the stick (which keeps that end of the stick in a fixed position). The lollipop's weight has magnitude $mg$, we are asked to find the normal force, and we don't know the tension. To keep this unknown tension from appearing in our upcoming $\sum\vect{\tau}$ expression, place the origin point $O$ at the lower left end of the stick so that tension has no moment arm. Point $O$ is fixed, so $\ddot{r}_O=0$ and $\sum\vect{\tau}=d\vec{L}/dt$.

Next, we choose a coordinate system centered on the origin point. It is convenient (but certainly not necessary!) to measure $\vec{L}$ in a coordinate system rotating with the lollipop. As shown in Figure \ref{fig:09_53}, we choose to let $\hat{x}$ be directed along the stick with $\hat{y}$ into the page and $\hat{z}=\hat{x}\times\hat{y}$. These unit vectors rotate with the lollipop, so they are not constant but vary with time. Because the time derivative of a rotating vector is its angular velocity crossed into the vector, the time derivatives of $\hat{x}$, $\hat{y}$, and $\hat{z}$ are
\begin{displaymath}
{\begin{array}{ccccl}
\dot{\hat{x}} & = & \vect{\Omega}\times\hat{x} & = & \Omega \cos\theta\hat{y} \\
\dot{\hat{y}} & = & \vect{\Omega}\times\hat{y} & = & -\Omega \cos\theta\hat{x}+\Omega\sin\theta\hat{z} \\
\dot{\hat{z}} & = & \vect{\Omega}\times\hat{z} & = & -\Omega \sin\theta\hat{y}.
\end{array} }
\end{displaymath}

\vspace{0.4cm}
\noindent{{Evaluating $d\vec{L}/dt$:}}
\vspace{0.2cm}

The lollipop is in purely rotational motion so $\vec{L}$ = $\vec{L}_{\mathrm{rot}} = \vec{I}\cdot\vect{\Omega}$ where $\vec{I}$ is the inertia tensor in the $xyz$ coordinate system centered on $O$. Because of the solid sphere's symmetry, any axis through its $CM$ is a principal axis and
\begin{displaymath}
\vec{I}_{CM}=
\left(\begin{array}{ccc}
\frac{2}{5}mr^2 & 0 & 0 \\
0 & \frac{2}{5}mr^2 & 0 \\
0 & 0 & \frac{2}{5}mr^2
\end{array}\right).
\end{displaymath}
Employing the parallel-axis theorem, its inertia tensor about the $xyz$ coordinate system centered on $O$ is
\begin{displaymath}
\vec{I}=
\left(\begin{array}{ccc}
\frac{2}{5}mr^2 + md^2 & 0 & 0 \\
0 & \frac{2}{5}mr^2 & 0 \\
0 & 0 & \frac{2}{5}mr^2
\end{array}\right)
\end{displaymath}
where $d^2=R^2+r^2$. Projecting $\vect{\Omega}$ onto the $xyz$ coordinate system gives $\vect{\Omega}=\Omega\sin\theta\hat{x}+\Omega\cos\theta\hat{z}$ so that 
\begin{equation}
\vec{L}=\vec{I}\cdot\vect{\Omega}=\left[\left(\frac{2}{5}r^2 + d^2\right)\Omega\sin\theta\right]\hat{x}+\frac{2}{5}mr^2\cos\theta\hat{z}.
\label{eq:L}
\end{equation}

Next, to calculate $d\vec{L}/dt$, notice in equation \ref{eq:L} that $\hat{x}$ and $\hat{z}$ vary with time while all other quantities are constant. Substituting the unit vectors' time derivatives from above, a line or two of algebra gives
\begin{displaymath}
\frac{d\vec{L}}{dt}=md^2\Omega^2\sin\theta\cos\theta\hat{y}.
\end{displaymath}
Substituting $\sin\theta=r/d$ and $\cos\theta=R/d$ from the geometry in Figure \ref{fig:09_53} gives
\begin{equation}
\frac{d\vec{L}}{dt}=m\Omega^2rR\hat{y},
\label{eq:dLdt}
\end{equation}
the quantity that, from $\sum\vect{\tau}=d\vec{L}/dt$, equals the net torque on the lollipop.

\vspace{0.4cm}
\noindent{{Applying $\sum\vect{\tau}=d\vec{L}/dt$ and solving:}}
\vspace{0.2cm}

The net torque about $O$ on the lollipop is due to the normal force and the weight, each of which has a moment arm of length $R$ (recall from above that tension on the massless stick exerts no torque about $O$) so that summing $\vect{\tau}=\vec{r}\times\vec{F}$ gives
\begin{equation}
\sum\vect{\tau}=\left(NR-mgR\right)\hat{y}
\label{eq:sumt}
\end{equation}
where $N$ is normal force and $g$ is the gravitational acceleration. 

Equating equation \ref{eq:dLdt} to equation \ref{eq:sumt}, canceling the common $R$ from each term, and solving for the normal force $N$ gives
\begin{displaymath}
\boxed{N=mg+mr\Omega^2}
\end{displaymath}
as required. This is an interesting result. The normal force on the lollipop is, perhaps surprisingly, independent of the angle $\theta$ and the distance $R$ from the fixed end to the point of ground contact. It counters not only the lollipop's weight, but also an additional force that increases with the size and mass of the lollipop and with the square of its rotation rate. This additional force is, of course, the downward vertical component of tension. The tension force, which did not appear in this solution because of our choice of origin, is responsible for holding the stick's lower left end in place and for providing the lollipop's centripetal acceleration.


\section{Conclusion}

Angular momentum is an interesting quantity and hopefully appears on students' lists of topics to learn more about as they progress from introductory, freshman-level physics into their upper-level courses. Having taught both freshman- and upper-level mechanics courses many times, I find that angular momentum is a rather difficult topic for students to master. The difficulties are both conceptual and mathematical, and I find the inherent conceptual difficulties are often compounded by oversimplified freshman-level presentations. I hope this paper helps to counter those misconceptions and reminds students of these points:
\begin{itemize}
\item Angular momentum is {\em not} generally parallel to angular velocity. Angular momentum is the moment of momentum and is in the direction obtained through $d\vec{L}=\vec{r}\times\vec{v}\,dm$.
\item An object does {\em not} need to be rotating to have angular momentum. Angular momentum is the moment of momentum and equals zero only when the integral of $d\vec{L}=\vec{r}\times\vec{v}\,dm$ over an object equals zero.
\item Along with many other relations, $\vec{L}=I\vect{\omega}$, $\sum\vect{\tau}=I\vect{\alpha}$, and $\sum\vect{\tau}=d\vec{L}/dt$ are {\em special cases}, only true under certain conditions. Physics students must be ever-watchful for these special cases, lest they apply them in error when the conditions do not hold.
\end{itemize}

\vspace{0.2cm}
Classical mechanics is a fascinating subject and I hope all freshman and sophomore physics students look forward to studying it in their upper-level courses.


\begin{acknowledgments}
J.~M.~Hughes gratefully acknowledges helpful discussions with B.~Berhane and M.~A.~Reynolds, each of whom patiently answered many questions from their next-office-door neighbor.
\end{acknowledgments}

\vfill

\end{article}


\begin{thebibliography}{}

\providecommand{\natexlab}[1]{#1}
\expandafter\ifx\csname urlstyle\endcsname\relax
  \providecommand{\doi}[1]{doi:\discretionary{}{}{}#1}\else
  \providecommand{\doi}{doi:\discretionary{}{}{}\begingroup
  \urlstyle{rm}\Url}\fi

\bibitem[{\textit{Morin}(2007)}]{Morin}
Morin,~D. (2007), \textit{Introduction to Classical Mechanics}, Cambridge University Press, Cambridge, UK.

\end{thebibliography}
\end{document}